**Generative AI as a Design Variable: An Evidence-Centered Framework for Principled Governance in STEM Assessment**

Yizhu Gao[1*], Zhongzhou Chen[2], Min Li[3], Xiaoming Zhai[1]

[1] University of Georgia, yizhu.gao@uga.edu; xiaoming.zhai@uga.edu
[2] University of Central Florida, zhongzhou.chen@ucf.edu
[3] University of Washington, minli@uw.edu
[*] Corresponding author: yizhu.gao@uga.edu

**Abstract**
*Generative Artificial Intelligence (GenAI) presents a governance challenge for STEM assessment. Unrestricted GenAI access enables task outsourcing that undermines the validity of traditional assessments; blanket prohibitions are difficult to enforce, may push use underground, and do little to prepare students for workplaces where GenAI-supported workflows are increasingly common. This paper addresses this dilemma by proposing a framework grounded in Evidence-Centered Design (ECD) that treats GenAI as a design variable within the assessment argument rather than an external threat to it. The framework analyzes how GenAI reshapes the student model, evidence model, and task model, and uses this analysis to articulate three principled governance stances. Restrict is warranted when GenAI would contaminate the inferential link between student work products and targeted unaided proficiency. Scaffold is warranted when bounded GenAI support can support peripheral demands without revealing the target construct, preserving inferential interpretability. Require is warranted when the target construct is disciplinary AI interaction competency and tasks can be designed to elicit process artifacts, including prompts, critiques, and revisions, that make student reasoning observable, scorable, and distinguishable from AI-generated output. This framework thus specifies when to restrict, scaffold, or require GenAI use in STEM assessment. To illustrate the Require stance, we present two task designs deployed in an introductory physics course and demonstrate that disciplinary AI interaction competencies are observable in student response artifacts and can be scored using defensible rubrics grounded in student data and expert knowledge. By situating GenAI governance within validity arguments, the framework offers actionable guidance for preserving learning integrity while supporting authentic preparation for AI-enabled professional environments.*

**Generative AI as a Design Variable in STEM Assessment: An Evidence-Centered Framework for Principled Governance**

## 1. Introduction

Generative Artificial Intelligence (GenAI) has prompted ongoing debate about its role in STEM assessment. On one hand, GenAI enables answer outsourcing by producing solutions that students may copy and submit, potentially diminishing the productive struggle that supports learning (Gao et al., 2025). On the other hand, GenAI can support students in engaging with complex, authentic problems and in practicing AI-mediated tasks that reflect emerging workplace demands (Otto et al., 2025). Preserving assessment validity while supporting authentic AI-mediated practice requires principled approaches to GenAI governance—specifying when its use should be restricted, scaffolded, or required.

Recent work has begun to address this need through governance frameworks for educational assessment. For example, Furze et al. (2024) proposed the AI Assessment Scale, a five-level scheme ranging from "no AI" to "full AI," which serves as a communication tool for defining permissible AI use. Building on this work, Perkins et al. (2024) refined these levels to no AI, AI planning, AI collaboration, full AI, and AI exploration, illustrating the diverse roles AI can play in assessment. Most recently, Chase and Galvin (2026) extended this line of work by arguing that decisions about permissible AI use should be anchored in learning objectives and the cognitive demands of assessment tasks. Together, these frameworks represent meaningful progress in making AI governance decisions explicit and communicable. Yet despite this progress, a fundamental validity question remains underexplored: to what extent does the integration of GenAI into assessment contexts alter the construct being measured, and to what extent the inferences drawn from student performance retain their validity when GenAI mediates the assessment process. Without a principled account of how GenAI reshapes the constructs being assessed, the evidence required to support those inferences, and the tasks needed to elicit that evidence, educators lack a validity-grounded basis for determining what forms of GenAI use are defensible in a given assessment. For instance, while "AI collaboration" has been characterized as involving students' critical evaluation and revision of AI-generated content (Perkins et al., 2024), the field has yet to establish whether AI collaboration constitutes a coherent and assessable target construct, what task designs can reliably elicit it, and what observable behavioral indicators serve as valid evidence of its presence.

To address these gaps, this study proposes an analytical framework that conceptualizes GenAI as a design variable within the assessment argument, examining how GenAI integration reshapes (a) the student model, (b) the evidence model, and (c) the task model. This framework is intended to provide validity-grounded guidance for decisions about when to require, scaffold, or restrict GenAI use in assessment. The study then illustrates how this framework can be operationalized in introductory physics through two Require-GenAI task designs and demonstrates that disciplinary AI interaction competencies are observable in student response artifacts and can be scored using rubrics grounded in both student performance data and expert knowledge.

## 2. Evidence-Centered Design as a Lens for Generative AI Governance

ECD is an approach to constructing educational assessments as evidentiary arguments (Mislevy et al., 2003). It begins with explicit claims about what learners should know and do (the student model), specifies the evidence needed to support those claims (the evidence model), and then designs tasks and scoring rubrics to elicit and evaluate that evidence under interpretable conditions (the task model). Together, these three models form an inferential chain linking task design to observable student performance to construct-level claims about learning. The explicit inferential structure of ECD makes it particularly well suited for analyzing how GenAI affects assessment validity: by specifying where and how inferences about student learning are grounded, it provides a principled basis for identifying where GenAI disrupts existing assessment arguments and where it might be leveraged to strengthen them. Each of the three models is introduced below, followed by an analysis of how GenAI integration could reshape it.

### *2.1. Student Model and Impacts of GenAI Use*

The student model defines the latent proficiency space targeted by an assessment (e.g. knowledge, skills, strategies) and specifies how these proficiencies are structured and related (e.g., skill dependencies, multidimensionality) (Mislevy et al., 2003). In STEM settings, the student model is typically anchored in established cognitive and performance frameworks. Common anchors include Bloom's taxonomy for characterizing cognitive demands (Zollman, 2012), national framework documents and standards that define disciplinary learning goals (Hoeg & Bencze, 2017), and career-readiness or 21st-century skills frameworks that emphasize transferable practices (Kennedy & Sundberg, 2020).

Historically, the constructs specified in the student model have evolved as new tools and social practices reshape what counts as target competence (Bennett, 2015). In STEM education, for example, the expansion of computer-based labs, simulations, and data-logging tools shifted the definition of competence from producing correct answers toward designing investigations and interpreting digitally generated evidence. GenAI is catalyzing a similar shift: students increasingly use GenAI to generate, critique, and refine ideas in course-related work, prompting renewed attention to constructs such as disciplinary AI literacy as an emerging component of contemporary disciplinary competence (Dera, 2025; Zhai, 2025; Zhao et al., 2024). Schleiss et al. (2024) delineate AI education into three categories: general AI literacy for the broad public, domain-specific AI literacy for disciplinary contexts, and expert-level AI literacy for computer science specialists. For STEM learners, disciplinary AI literacy is particularly relevant. Zhai (2025) defines it as "the integrated capacity to understand, apply, and critically reflect on AI in authentic disciplinary practices", operationalized through six components: conceptual knowledge, practical tool use, critical evaluation, collaborative and reflective disposition, data and computational literacy, and ethical and societal understanding. Complementing this, Dera (2025) highlights a gap in research on discipline-specific AI use—specifically, the lack of assessments that measure students' demonstrated performance of responsible AI use rather than their self-reported perceptions.

Addressing this gap, Li et al. (2025) move beyond conceptual definitions of AI literacy to propose a behaviorally grounded, performance-based rubric framework for assessing how students actually interact with GenAI. Rather than inferring competence from self-report, their framework captures observable interaction behaviors produced during authentic tasks, organized around three dimensions. First, *Information Retrieval and Analysis* captures how students structure and refine their queries to GenAI, operationalized through six indicators: the number and cognitive level of prompts, whether students prompt from multiple perspectives, whether they follow up based on AI responses, whether they continuously refine prompts, and whether their prompt chain exhibits logical structure. Second, *Information Evaluation* captures how students critically assess AI-generated responses, operationalized through three indicators: the frequency and quality of justified queries, the degree to which students screen and select relevant evidence, and whether they critically evaluate the relevance, credibility, and logical strength of AI outputs. Third, *Information Integration* captures how students synthesize AI-generated information into coherent arguments, operationalized through two indicators assessing whether students justify conclusions from multiple perspectives and whether their argumentation is logically consistent rather than a direct copy of AI responses. Together, these three dimensions offer a more operationalizable basis for specifying the student model in assessment contexts where GenAI is an explicit part of the task—moving beyond latent competency descriptions toward observable interaction behaviors that can be directly linked to student response artifacts. These behaviors, in turn, provide the empirical foundation for the evidence and task models discussed in the sections that follow

### *2.2. Evidence Model and Impacts of GenAI Use*

Evidence models specify how beliefs about student model variables are updated based on observed performance captured in learners' work products from assessment tasks (Mislevy et al., 2003). An evidence model comprises two complementary components: evidence rules and a measurement model. Evidence rules define the procedures for extracting observable variables from what students produce in a task. The measurement model then links these observable indicators to the targeted student model variables, specifying how patterns of evidence raise or lower the inferred likelihood that a learner possesses particular knowledge, skills, or misunderstandings.

GenAI complicates the evidence model by undermining evidence rules—the procedures that map observed student work products to latent constructs—because when GenAI contributes to those products, it becomes unclear whether they reflect student competence or AI capability. Consider an evidence rule that operationalizes the correctness of a physics response as a binary score: that score is no longer interpretable as a direct indicator of the targeted construct if the final solution was generated or substantially shaped by GenAI. The inferential link from product features—such as correctness or explanation quality—to student proficiency breaks down, because high-quality outputs can be outsourced to GenAI without engaging the competencies the assessment is designed to measure.

This threat to inferential validity is not entirely novel. Outsourcing of STEM problem solving has long challenged assessment integrity (e.g., via homework-help platforms)—but GenAI intensifies the problem qualitatively, rather than merely in degree. Contemporary LLMs can solve many conventional textbook-style items at passing or above-passing levels: Kortemeyer (2023) found that ChatGPT could narrowly pass a calculus-based introductory physics course when graded on representative course assessments. Recent systems have also demonstrated near–gold-medal performance on Olympiad-level mathematics under controlled evaluation settings (Jian et al., 2025). However, this disruption is not uniform: for tasks requiring factual recall or standard algorithmic problem solving, GenAI can substitute for student cognition almost completely; for tasks requiring disciplinary judgement applied to novel or ambiguous problems, its substitution capacity is more limited, preserving spaces where student knowledge must be actively deployed. This asymmetry suggests that the validity of evidence rules under GenAI depends critically on the cognitive demands of the task (Gao et al., 2025). This shifts the focal observation from products that GenAI can readily generate toward behaviors and artifacts that are more resistant to outsourcing, and from what answer was produced to how disciplinary knowledge was deployed in producing it.

### *2.3. Task Model and Impacts of GenAI Use*

Task models describe how to design assessment situations that will elicit the evidence needed for the evidence models (Mislevy et al., 2003). A task model defines the key features of a task—such as the materials presented to the learners and the types of work products the learners generate in response—and serves as a template for a family of tasks that share the same evidentiary purpose. Specific tasks are created by instantiating the model: selecting or authoring the presentation materials and assigning values to the task-model variables that control task features and difficulty.

GenAI reshapes what assessment tasks are viable and how they should be designed. By enabling generic systems to solve many traditional text-based, static problems, it increases the risk of outsourcing and weakens such items' validity as measures of individual student thinking. This shift motivates task designs that are richer and more contextualized, and that treat GenAI as an explicit element of the prompt, interaction, or toolset. Although higher-order constructs are most appropriately assessed through open-ended, authentic performances (Wiggins, 1991), assessment systems have historically relied on closed-ended proxies because complex performances can introduce construct-irrelevant variance and reduce interpretability (Messick, 2013). GenAI can help relax this constraint by scaffolding multi-step work while systematically eliciting and logging intermediate products that provide targeted evidence aligned with intended claims.

Two emerging task design approaches illustrate this expanded design space. Conversation-based assessment structures the assessment as an iterative dialogue, for example, requiring students to construct a scientific argument, respond to AI-generated counterevidence, and revise their claims accordingly, generating a natural log of process artifacts that renders disciplinary reasoning visible at multiple decision points rather than only in a final product (Zapata-Rivera et al., 2026). Each exchange in a dialogue constitutes an observation opportunity, requiring students to deploy disciplinary knowledge in real time rather than submitting a single polished product. Related approaches employing constrained AI support, in which the system provides bounded hints or feedback that scaffold peripheral skills without disclosing target solutions, similarly preserve the evidentiary integrity of student responses (Kapoor et al., 2025; Liffiton et al., 2023). In this regime, GenAI functions like a calculator in mathematical proof or a reference manual in experimental design: it reduces peripheral cognitive load while students continue to demonstrate the core disciplinary reasoning the assessment is designed to measure. This expanded design space is particularly consequential in STEM, where task designs can position students to engage with authentic scientific and engineering practices, constructing arguments, designing investigations, and evaluating competing models, rather than relying on simplified proxies that reduce these practices to decontextualized written responses.

## 3. Positioning Generative Artificial Intelligence in STEM Assessment

The preceding analysis reveals that GenAI's impact on assessment validity is not uniform: it depends on what construct is being targeted, what evidence is being used to support the inference, and how the task is designed to elicit the evidence. These three dependencies suggest a structured decision process for GenAI governance. Building on this analysis, we propose an ECD-based decision framework to guide assessment designers and instructors in determining the appropriate GenAI governance mode for a given assessment context (see Figure 1).

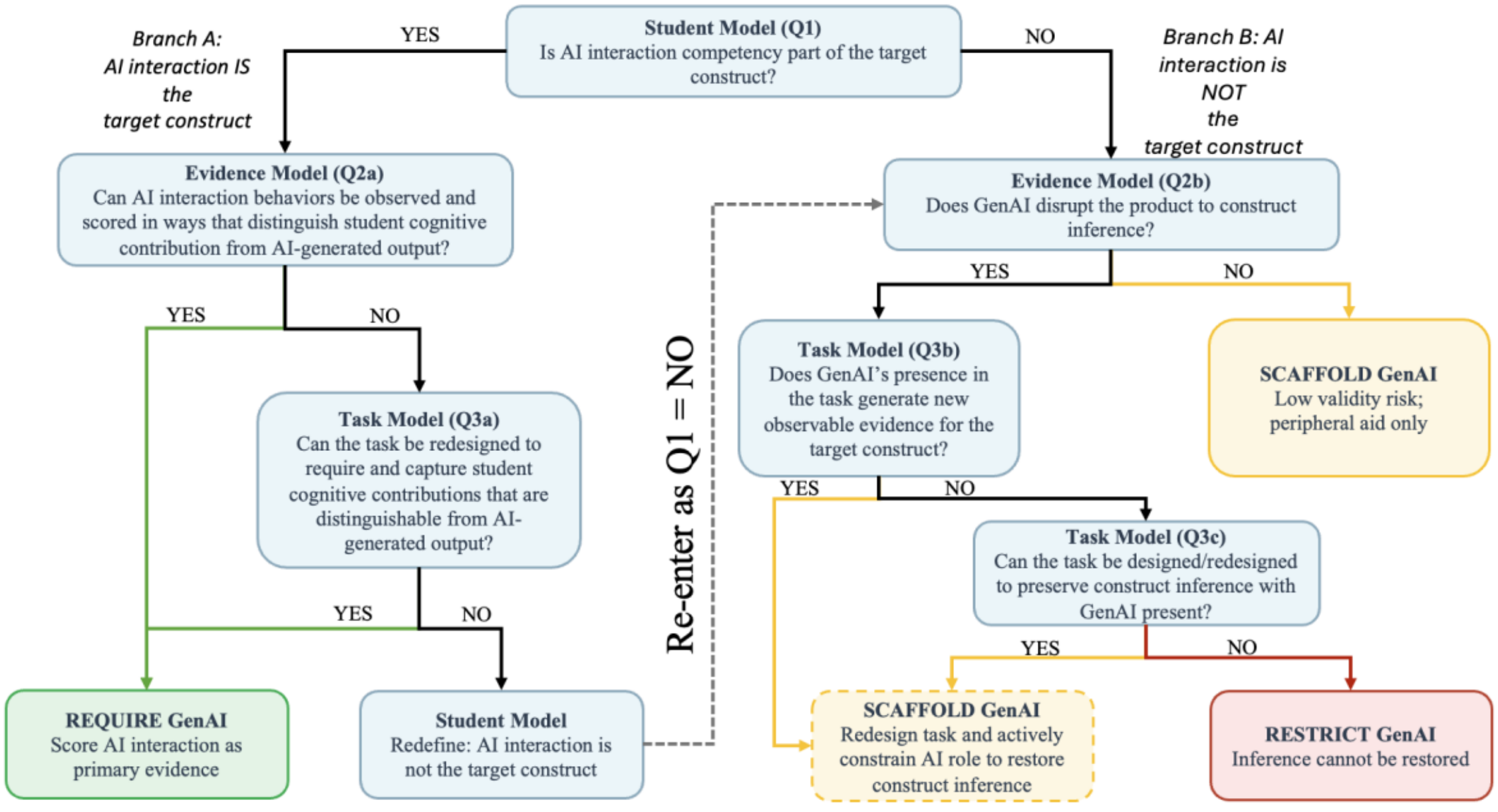

***Figure 1. GenAI Governance in STEM Assessment***

*Note.* The framework applies the three ECD models as sequential decision criteria for determining GenAI governance mode across two branches: Branch A, in which AI interaction competency is the target construct, and Branch B, in which it is not. Blue boxes represent ECD decision nodes; green boxes indicate a Require governance decision; yellow boxes with solid borders indicate a Scaffold (redesign) decision; yellow boxes with dashed borders indicate a Scaffold (low validity risk) decision; and red boxes indicate a Restrict decision. The grey dashed path represents a construct revision loop, in which the assessor redefines the target construct and re-enters the framework at Branch B. Governance decisions are not static; as GenAI capabilities and task designs evolve, answers to each question may shift, requiring ongoing reassessment.

### *3.1. Require GenAI*

A Require governance mode is warranted when AI interaction competency is itself the target construct—that is, when the assessment is explicitly designed to measure how students engage with GenAI as a disciplinary practice rather than treating AI use as prohibited. In this mode, GenAI is not a validity threat but a measurement requirement: the student model specifies AI-mediated disciplinary reasoning as the targeted proficiency, the evidence model recovers AI interaction behaviors recovered from direct process records or student self-reports in ways that distinguish student cognitive contribution from AI-generated output, and the task model designs assessment situations that elicit those behaviors. Operationally, this requires tasks that prompt students to engage in the cycle of AI-mediated reasoning— iteratively generating queries and critically evaluating AI outputs against disciplinary standards (Li et al., 2025).

### *3.2. Scaffold GenAI*

A Scaffold governance mode is appropriate under three distinct conditions that share a common underlying logic: the core disciplinary construct remains the assessment target, and GenAI is permitted only insofar as it does not substitute for the reasoning the assessment is designed to measure.

The first condition arises when GenAI disruption of the evidence model is judged to be minimal (Q2b = NO), typically because the task's cognitive demands substantially exceed GenAI's current substitution capacity, as in open-ended experimental design tasks that require disciplinary judgment about course-specific constraints and theoretical framing that GenAI cannot reliably supply. In this condition, GenAI may be permitted as a peripheral aid—for example, to retrieve equipment specifications or recall formula names—without meaningfully compromising the evidentiary chain. No task redesign is required; the existing task structure naturally limits GenAI's substitution capacity, and instructors do not need to specify or enforce boundaries on how GenAI may be used.

The second condition arises when GenAI disruption is substantial (Q2b = YES) but GenAI's presence in the task generates new observable evidence for the target construct (Q3b = YES)—that is, when the task is designed so that student-AI interaction makes disciplinary reasoning visible at multiple decision points rather than only in a final product. Conversation-based assessment formats exemplify this condition across a range of construct complexity. For complex constructs such as scientific argumentation, conversation-based assessment requires students to construct claims, respond to AI-generated counterevidence, and revise their reasoning across multiple exchanges, generating a natural log of process artifacts that renders disciplinary reasoning visible throughout the assessment episode rather than only at its conclusion (Zapata-Rivera et al., 2023). For simpler constructs, such as conceptual understanding of a single principle or procedural application of a formula, conversation-based assessment similarly offers evidentiary advantages: by prompting students to explain their reasoning at each step rather than submitting a final answer, the format makes visible the thinking behind the product, allowing raters to distinguish a student who arrived at a correct answer through sound disciplinary reasoning from one who did so through trial and error or surface-level pattern matching. In both cases, the key mechanism is the same: GenAI's role as interlocutor creates natural observation points that a static task cannot produce, making student reasoning recoverable from the artifact record rather than inferable only from the final product. In this condition, GenAI is not merely tolerated despite its disruptive presence but actively leveraged as an evidentiary resource—the task is designed around GenAI's generative capacity precisely because doing so produces richer evidence of student reasoning than a non-AI task could elicit.

The third condition arises when GenAI disruption is substantial (Q2b = YES), GenAI's presence does not generate new construct evidence (Q3b = NO), but task redesign can restore evidentiary integrity (Q3c = YES). Unlike the second condition, where GenAI is leveraged as an evidentiary asset, this condition treats GenAI as a design constraint to be guardrailed: the goal of redesign is not to leverage what GenAI generates but to bound what GenAI is permitted to do, ensuring that the core disciplinary reasoning targeted by the assessment remains the student's responsibility. This approach is analogous to permitting a calculator in a physics examination—the tool handles peripheral computation while the student continues to demonstrate the conceptual and analytical reasoning the assessment is designed to measure. Concretely, guardrailing GenAI means specifying the functions GenAI may perform and the functions it may not. For example, in a physics problem solving task, GenAI might be permitted to retrieve relevant formulas or explain concepts but prohibited from generating or checking numerical solutions—preserving the student's responsibility for the core quantitative reasoning the task targets.

### *3.3. Restrict GenAI*

A Restrict governance mode is warranted when GenAI substantially disrupts the evidence model (Q2b = YES), GenAI's presence in the task does not generate new observable evidence for the target construct (Q3b = NO), and no task redesign can restore the inferential chain from student work products to the targeted construct (Q3c = NO). This sequential condition is theoretically important: Restrict is not triggered by the mere presence of GenAI disruption, but only after both Q3b and Q3c have been answered in the negative—confirming that neither leveraging GenAI's generative capacity nor guardrailing GenAI within a redesigned task can preserve construct validity. The Q3c = NO condition specifically means that no guardrail configuration, no matter how carefully the permitted and prohibited functions of GenAI are specified, can prevent GenAI from substituting for the core reasoning the construct targets. When the construct itself requires unassisted individual performance, the guardrail boundary cannot be maintained, and restriction is the only validity-preserving option.

## 4. Illustrating the Framework: Require-GenAI Tasks in Physics Education

The preceding sections established a theoretical framework for GenAI governance in STEM assessment, grounding the three ECD models and operationalizing validity concerns into a sequential decision process organized across two branches. Among the three governance modes, Require-GenAI tasks represent the most novel and theoretically demanding case: they reposition GenAI from a threat to assessment validity into a constitutive element of the construct being measured. This section illustrates how Require-GenAI tasks can be designed and scored in physics, demonstrating how the framework's abstract commitments translate into concrete task design decisions, evidence rules, and scoring rubrics.

### *4.1. Task Design*

To illustrate how Require-GenAI governance mode is operationalized in practice, this section presents two tasks deployed in an undergraduate-level introductory physics course. The first task "*Remora Free Body Diagram*" (see Figure 2) presents students with an image of a remora fish attached to a shark and asks them to collaborate with GenAI to construct the most accurate free body diagram possible. The domain problem meets the accuracy-ceiling criterion because an accurate free body diagram requires identifying forces that are specific to remora biology—including the suction force generated by the modified dorsal fin acting as a disc, hydrodynamic drag, and the normal force from the host's surface—forces that GenAI consistently failed to produce correctly without disciplinary guidance from the student. The task elicits five categories of process artifact through its structured prompt items: the GenAI tool and model used (Q1), a summary of the prompting conversation (Q2), the final GenAI output (Q3), the student's critique of errors or limitations in that output (Q4), and a reflection on the assumptions and simplifications GenAI introduced (Q5). While Q1 identifies the tool use, Q3 provides the AI output as a referential baseline, and Q5 elicits assumption reflection, Q2 and Q4 are the primary evidence resources for the two targeted dimensions: Q2 for Physics-Driven AI Interaction and Q4 for Physics-Grounded Evaluation. Together, these two artifacts make student cognitive contribution observable and distinguishable from AI-generated content, because they require students to document their own reasoning about what GenAI got wrong and why. Students who accept GenAI output uncritically are detectable because their FBDs will reflect generic force configurations rather than the biologically specific forces the task demands.

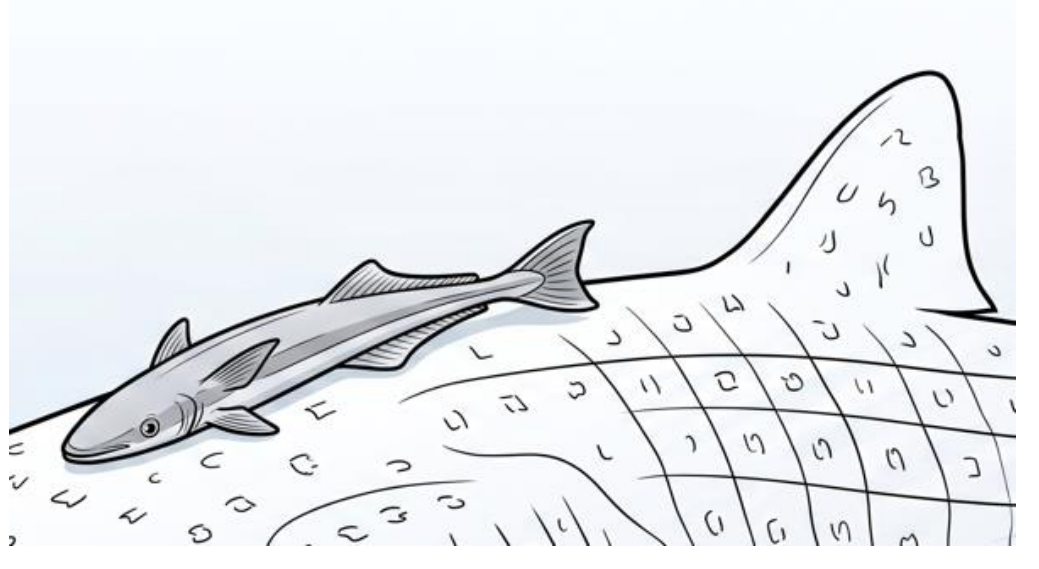

The fish in this image is called a remora, or a sucker fish. Your task is to collaborate with Generative AI and create the best free body diagram for the remora fish in the image. Tip: Ask GenAI for any information you need to learn about the remora fish and the sucker plate. **In your response, answer the following questions:**

1. Which Generative AI tools and services did you use to answer the question, and on which day (because GenAI services can update rapidly). For example: ChatGPT 5 Thinking 09/03/2025. Google Gemini 2.5 pro 08/01/2025
2. Summarize the conversation(s) you had with GenAI, outlining what instructions and information did you give GenAI for it to reach the final output?
3. Overall, how many attempts did it take for the GenAI response to generate the answers you need? If more than once, show only the final output or the best output.
4. If the GenAI response was imperfect or imperfect, list everything that the GenAI output got wrong or can be improved.
5. Ask GenAI to list all the assumptions and simplifications that went into creating the results. Did GenAI make any important assumptions or simplifications?

***Figure 2. Require-GenAI Task "Remora Free Body Diagram"***

*Note.* The image above is a recreation of an actual Remora fish, used for illustrative purposes.

The second task "*Water Balloon Physics Image*" (see Figure 3) assigns GenAI a dual role: it is simultaneously the source of the artifact being analyzed and the tool students use to correct it. Students receive a GenAI-generated image of a child swinging a water balloon on a massless string in a horizontal circle and are asked to identify physically impossible or implausible elements, use GenAI to generate a more accurate image, and ask GenAI to construct a correct free body diagram of the balloon. The domain problem meets the accuracy-ceiling criterion because correctly identifying violations of Newtonian mechanics in the image—such as an implausibly taut string at an incorrect angle, absent tension forces, or misrepresented centripetal acceleration—requires the student to bring their own understanding of circular motion to bear, and because GenAI consistently missed at least one expert-anchored error in the image and FBD, as confirmed by the expert answer key. The three-way error classification structure in Part 1 (what GenAI identified, what the student identified independently, and what the student believes GenAI misidentified) is particularly important for the evidence model: it directly operationalizes the distinguishability condition of Q2a by making the student's independent physical reasoning directly observable and separable from GenAI's contribution. A student who cannot independently identify errors beyond what GenAI flags produces an artifact that fails the distinguishability condition, making uncritical acceptance detectable in the same way as in the Remora task.

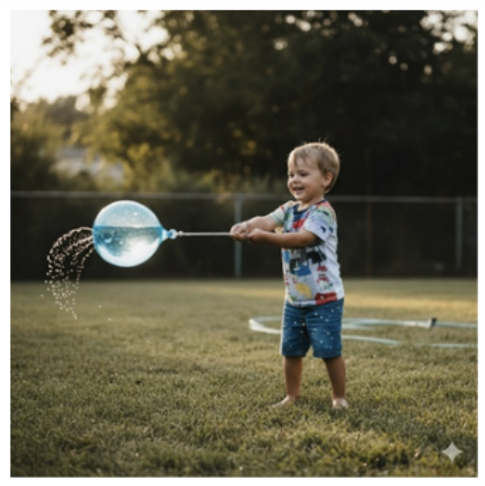

Here is an image of a small child holding a water balloon with a long massless string and swinging it in a horizontal circle. This image is generated by a Generative AI service using a simple prompt. Your task is as follows:

1. Examine this image and identify everything that is physically impossible or highly unlikely about this image, using the knowledge you've learned in this class. If you used Generative AI to assist you with identifying the incorrect things, list the following:
   - things that GenAI identified as impossible
   - Things that GenAI missed, and you identified as impossible. Give justification as to why it is impossible.
   - Things that GenAI identified as impossible, but you think GenAI is wrong (if any).
2. Upload this image into generative AI and ask it to create a more physically accurate image, by telling it what was wrong with the initial image. List both your prompt and the final image outcome below.

Note: If you did multiple trials, summarize the key aspects of your entire conversation and paste the final image. If the final image still isn't completely correct, list in what way it is not perfect.

3. Ask GenAI to draw the correct free body diagram (FBD) of the water balloon being swung in a horizontal circle. Give it enough information so that the free body diagram is as accurate as possible. Include the following information in your answer:
   1. Your initial prompt to GenAI
   2. The initial AI generated FBD
   3. (If initial FBD is incorrect) A summary of any follow up prompts that you gave the AI to improve the initial FBD
   4. (If initial FBD is incorrect) The final FBD and anything that is still incorrect about the final FBD.

Note: Not all Generative AI services are able to create images. Use one that is able to do so. Which AI did you use to solve this problem (Gemini, Copilot, Claude, etc.), and what thinking mode did you use (such as Fast, Thinking, Pro, etc.)?

***Figure 3. Require-GenAI Task "Water Balloon"***

Both tasks position GenAI as a constitutive element of the assessment rather than an external threat to it, and both require students to produce layered evidence—prompts, critiques, revised artifacts—that reflects the quality of their disciplinary reasoning about GenAI outputs.  Both tasks satisfy the conditions for Require governance: Q1 is met because AI interaction competency is the target construct, and Q2a is met because the process artifacts produced make student cognitive contributions observable, scorable, and distinguishable from AI-generated output.

### *4.2. Constructs, Evidence Rules, and Scoring Rubrics*

**Constructs.** Drawing on Li et al.'s (2025) HAII-Capability framework, which defines human-AI information interaction competency across four components—Information Retrieval, Information Analysis, Information Evaluation, and Information Integration—two dimensions were identified as directly observable from the task artifacts. The first dimension, Physics-Driven AI Interaction (D1), integrates the Information Retrieval and Information Analysis components: it captures the degree to which students use disciplinary physics knowledge to structure the initial prompt and adaptively drive subsequent interaction—distinguishing students who actively frame and refine the problem using physics knowledge from those who passively accept AI output without disciplinary engagement. The second dimension, Physics-Grounded Evaluation (D2), corresponds to the Information Evaluation component: it captures the degree to which students critically assess AI-generated physics content against disciplinary standards—distinguishing students who identify and justify physics-grounded errors or omissions from those who accept AI output as authoritative. The Information Integration component is reflected implicitly in the highest performance levels of both dimensions rather than scored independently, because the tasks elicit integration through the FBD construction and critique process rather than as a separable artifact.

**Evidence rules.** Evidence rules specify the procedures for extracting observable indicators from student artifacts and linking them to the targeted dimensions. For the Remora task, the primary evidence source for *Physics-Driven AI Interaction* is Q2 (prompt summary). Because the Remora task does not require students to submit interaction logs directly, Q2 functions as a self-report of the prompting conversation. The rater uses Q2 to recover two things: what physical information the student supplied to GenAI in the initial prompt, and whether the student re-prompted after judging the output insufficient. Three evidence sources are used jointly for *Physics-Grounded Evaluation*. Q3 (GenAI output) serves as the referential baseline, allowing the rater to contextualize the student's critique by seeing what AI actually produced, and to assess whether the student's identified errors are genuine physics-level problems or surface misreadings. Q4 (critique of AI output) provides the primary evidence for error identification and physical justification: the rater assesses whether errors are identified at the surface, physics, or modeling level, and whether each identified error is accompanied by a physics-based reason connecting it to a disciplinary principle. Q5 (assumption reflection) provides supplementary evidence for the independence of the student's evaluative judgment: the rater assesses whether the student's assumption list goes beyond what AI provided, reflecting physical reasoning about the specific scenario rather than a paraphrase of AI output.

For the Water Balloon task, the primary evidence sources for *Physics-Driven AI Interaction* are the Part 2 prompt (instruction to generate a more physically accurate image) and the Part 3 initial prompt (instruction to construct a correct FBD), supplemented by the Part 3 follow-up summary where applicable. The rater assesses two things from these sources: whether the student's initial prompts specify the physical errors or conditions that need to be corrected—indicating that the student brought their own physical knowledge to bear before seeing AI output—and whether follow-up prompts in Part 3 are driven by physics judgment about what the initial AI output got wrong. For *Physics-Grounded Evaluation*, three evidence sources are used jointly. Part 1 (three-way error classification) provides the primary evidence for the independence and accuracy of the student's error identification. The three-way structure—separately listing what GenAI identified, what the student identified independently, and what GenAI got wrong— directly operationalizes the distinguishability condition: by requiring students to attribute each error to a source, the task design makes independent judgment directly observable and separable from AI contribution. The rater focuses particularly on the second category (errors the student identified independently) and the third category (errors the student believes GenAI misidentified), as these are the cells in which independent disciplinary judgment is most directly evidenced. Part 2 (revised image and prompt) provides supplementary evidence for evaluation quality: the physical corrections the student requested from GenAI reveal what errors the student judged to be most important, and whether those judgments are grounded in Newtonian mechanics. Part 3 (FBD critique) provides additional evidence for *Physics-Grounded Evaluation*, particularly for modeling-level evaluation: the student's assessment of what remains incorrect in the final AI-generated FBD—including identification of spurious forces such as centripetal force added as a separate vector—requires the student to evaluate AI output against a disciplinary model of circular motion rather than surface features of the diagram. The expert answer key—which identifies string angle, balloon deformation, and spurious centripetal force addition as key errors—serves as the external calibration standard for the highest *Physics-Grounded Evaluation* performance levels across both Part 1 and Part 3.

In both tasks, the expert answer key serves as the external calibration standard for the highest *Physics-Grounded Evaluation* performance levels. For the Remora task, the expert identified friction direction and host acceleration assumptions as key errors commonly missed by both AI and students; for the Water Balloon task, the expert identified string angle, balloon deformation, and spurious centripetal force addition as key errors. A student whose critique independently identifies expert-anchored errors — without having been prompted to do so by AI output—provides the strongest evidence of physics-grounded evaluation. The expert answer key also empirically verifies the accuracy-ceiling criterion for both tasks: because GenAI consistently missed at least one expert-anchored error, both tasks confirm that student disciplinary input is necessary to reach the highest performance levels, satisfying the Require mode's accuracy-ceiling design principle.

**Scoring rubrics.** Four scoring rubrics were developed to operationalize the two targeted dimensions across the two tasks. For each task, one rubric captures Physics-Driven AI Interaction (D1) and one captures Physics-Grounded Evaluation (D2). Although the rubrics are task-specific in their examples and physical conditions, the underlying scoring logic is shared across both tasks within each dimension. The D1 rubric is a 0–4 scale that captures the degree to which students use disciplinary physics knowledge to structure and drive their AI interaction (see Tables 1 and 3). Scores are assigned based on two criteria evaluated jointly: the specificity of physical conditions present in the interaction, and whether physics-driven follow-up iterations are present. Crucially, the rubric evaluates the entire interaction rather than the initial prompt alone: a weak initial prompt followed by substantive physics-driven follow-ups can score as highly as a strong initial prompt that produced accurate AI output without requiring correction.

The D2 rubric is a 0–3 scale that captures the accuracy and depth of students' physical judgment about AI-generated content (see Tables 2 and 4). The scoring logic is anchored in the expert answer key: both the AI's output and the student's evaluation are assessed against expert-defined key errors and assumptions. For the Remora task, D2 is applied to Q3/Q4 (FBD critique) and Q5 (assumption reflection) separately, as these components target different aspects of evaluative judgment. For the Water Balloon task, D2 is applied to Part 1 (image analysis) and Part 3 (FBD critique) separately. In both tasks, the expert answer key serves as the external calibration standard: The highest score on D2 requires that the student independently identify at least one expert-anchored error with physical justification, without having been prompted to do so by AI output, reflecting the deepest level of physics-grounded evaluation.

**Table 1. Scoring Rubrics for Physics-Driven AI Interaction (D1): Remora Free Body Diagram Task**

| Indicator | Scoring Rubrics | | | | |
|---|---|---|---|---|---|
| | 0<br>No physics-driven interaction | 1<br>Minimal physics framing | 2<br>Adequate physics framing | 3<br>Proficient physics-driven interaction | 4<br>Advanced physics-driven interaction |
| Q2 from the Remora task | Example: "Create a free body diagram for the | Example: "Can you create a free body diagram of all the forces | Example: "I started by asking GenAI to analyze the image of the remora fish and to list all of the information needed to create a free | Example: "I asked: Create a free-body diagram of a remora fish suctioned to a larger fish | Example: "I asked the model to: (1) explain how a remora's suction disc adheres to a host, (2) list the dominant forces on a remora attached to a |
| | remora fish in the images.” | associated with a remora fish attached to a host that is swimming through the ocean.” | body diagram of the remora fish. I asked it to only take into account forces that are acting on the remora fish. It included equations on how to calculate the forces in the image and I asked it to remove the equations. I kept trying to fix the way the arrows in the image were pointing, and I even included an image for it to use as reference.” | (like a whale shark). I then specified the swimming direction, asked whether suction force points downward given the dorsal attachment, and clarified whether normal force and suction force are distinct forces and their directions.” | moving host in water, (3) sketch a clean free-body diagram with labeled forces and a simple coordinate frame, and (4) given equilibrium relations and standard formulas (drag, buoyancy, friction) appropriate for an introductory physics analysis. I also asked it to assume the remora is hitchhiking (no thrust) and the local water flow is roughly parallel to the host’s surface.” |

***Note.*** **No physics-driven interaction (0)**: Prompt contains no physics content; generic task request only with no reference to forces, motion, or physical conditions. No physics-driven follow-up at any point in the interaction. **Minimal physics framing (1)**: Physics topic mentioned in prompt (e.g., “forces,” “free body diagram,” “remora fish”) but without specific physical conditions or parameters; no physics-driven follow-up. Student showed basic awareness of the disciplinary context but did not use physics knowledge to direct the AI interaction. **Adequate physics framing (2)**: At least one specific physical condition appears somewhere in the interaction—either in the initial prompt or in a physics-driven follow-up. This can come from: (Path A) a strong initial prompt specifying at least one physical condition (e.g., a specific force type, motion state, or attachment geometry), with AI output sufficiently accurate that no corrective follow-up was needed; OR (Path B) a weak initial prompt followed by at least one physics-driven follow-up that directed the interaction toward disciplinary accuracy. Student demonstrated physics-grounded AI interaction through either initial framing or targeted iteration. **Proficient physics-driven interaction (3)**: Multiple specific physical conditions appear in the interaction AND at least one physics-driven follow-up is present; OR the initial prompt specifies multiple physical conditions and AI output was sufficiently accurate that no corrective follow-up was needed. Student actively controlled the interaction using disciplinary knowledge through structured multi-condition framing, with or without targeted iteration. **Advanced physics-driven interaction (4)**: Prompt structures the interaction around multiple disciplinary goals AND multiple physics-driven follow-up iterations are present. Student demonstrated both proactive disciplinary framing and sustained iterative refinement—the full cycle of physics-driven AI interaction.

## Table 2. Scoring Rubrics for Physics-Grounded Evaluation (D2): Remora Free Body Diagram Task

| Indicator | Scoring Rubrics<br>0<br>No evaluation or inaccurate evaluation | 1<br>Surface-level evaluation | 2<br>Physics-grounded evaluation | 3<br>Modeling-level evaluation |
|---|---|---|---|---|
| Q3, Q4 from the Remora task | Example: “Gemini’s outputs include errors (Q3). And student answered Q4: I did not have any incorrect responses from Gemini.” | Example: “Gemini’s outputs include errors (Q3). And student answered Q4: I believe the AI is imperfect with the fact that it labeled drag multiple times, where I believe it could have used one arrow to represent drag once.” | Example: “Q4: While ChatGPT didn’t get anything inherently wrong (as far as I know), I did notice that it referred to suction as an actual force at some points, while at other times rightfully referring to it as a downward net force from pressure effects.” | Example: “I think that Copilot got the suction force wrong because the remora is on the top of the fin, not on the bottom. Also, weight is stated correctly however it points in the wrong direction.” |
| Q5 from the Remora task | Example: “Gemini’s outputs missed assumptions. And student responded Q5: I did not observe any assumptions missed by GenAI.” | Example: “Internal forces such as muscle tension, fin movement, and metabolic activity are not included.” | Example: “The listing was pretty accurate from ChatGPT, there are some considerations though that I had thought about myself: The water in the ocean can change such as temperature, pressure gradients, or things like rip currents. Also, the drag force can also torque, and the remora can twist off similar to a bottle cap.” | Example: “Gemini’s outputs missed assumptions. And Q5: 1. Constant velocity (no acceleration). 2. Host surface locally flat; flow steady near remora. 3. Thrust = 0 if remora passively attached. 4. Ignore fluid turbulence, salinity/temperature effects.” |

***Note.*** **For Q3 and Q4: No evaluation or inaccurate evaluation (0):** Student provides no independent evaluation of the FBD; accepts AI output as correct without any physical assessment. OR student’s independent evaluation is inaccurate—accepts AI’s errors without questioning, or incorrectly challenges AI’s correct identification of expert key errors. No engagement with expert-defined key errors. **Surface-level evaluation (1):** Student notices an error in the FBD or agrees with AI’s analysis, but without physical justification. Includes: identifying a visual or labeling issue without connecting it to a physics principle; noting that a force appears multiple times without explaining why this violates the physics of the scenario. Student shows independent engagement but reasoning is not grounded in disciplinary principles. **Physics-grounded evaluation (2):** Student identifies at least one error in the FBD with physical justification, but the error identified is not among the expert-defined key errors. Evaluation is connected to at least one physics law or principle specific to the scenario. **Modeling-level evaluation (3):** Student identifies at least one expert-defined key error with physical justification. Justification references disciplinary principles and connects the identified error to a specific physics law or principle. **For Q5**: **No evaluation or inaccurate evaluation (0):** Student provides no independent evaluation of AI’s assumptions; accepts AI’s list as complete without any physical assessment. OR student’s evaluation is inaccurate—accepts AI’s incorrect assumptions, or fails to identify any expert-defined key assumptions that AI missed. No engagement with expert-defined key assumptions. **Surface-level evaluation (1):** Student identifies at least one assumption not present in AI's list, but without sufficient physical justification; OR the assumption identified is not among the expert-defined key assumptions and lacks grounding in the specific physical scenario. Student shows independent thinking beyond copying AI’s list but reasoning is not connected to disciplinary principles. **Physics-grounded evaluation (2):** Student identifies at least one assumption beyond AI’s list with physical justification, but the assumption identified is not among the expert-defined key assumptions. Evaluation is connected to at least one physics law or principle specific to the scenario. **Modeling-level evaluation (3):** Student identifies at least one expert-defined key assumption with physical justification. Justification references disciplinary principles and connects the identified assumption to a specific physics law or principle. Expert-defined key assumptions include: gravity direction assumes the photo is taken at a standard orientation; host fish is assumed to be moving at constant velocity (no acceleration or turning); remora is treated as a point particle; lift and fluid turbulence are ignored.

**Table 3. Scoring Rubrics for Physics-Driven AI Interaction (D1): Water Balloon Task**

| Indicator | Scoring Rubrics<br>0<br>No physics-driven interaction | <br>1<br>Minimal physics framing | <br>2<br>Adequate physics framing | <br>3<br>Proficient physics-driven interaction | <br>4<br>Advanced physics-driven interaction |
|---|---|---|---|---|---|
| Part 2 | Example: "Initial prompt: Examine this image and identify everything that is physically impossible or highly unlikely about this image' Follow-up: can you create a more physically accurate image. " | Example: "Initial prompt: Considering what makes the image physically impossible, generate a new, more physically accurate image. Follow-up: asked for the balloon to be "a bit deflated." | Example: "I uploaded the original image and asked the generative AI to make it more physically accurate. Specifically, I asked it to correct the direction of the water droplets so they move tangentially to the circular motion instead of downward, to make the water balloon appear heavier and more realistic in density, and to make the string look flexible rather than stiff. I tried to fix the direction of the water splashes, the visual density of the water balloon, and the flexibility of the string. However, not everything turned out perfectly. The density and weight of the balloon improved slightly, but I could not fully fix the string, it still appears somewhat stiff and overly tight instead of flexible and curved under tension." | Example: "create a more physically accurate image take into account that: according to the image, it is possible to highlight that the water is falling directly downward, completely ignoring the direction of movement, which in a real example, the water would follow the opposite direction of velocity. On the other hand, centrifugal forces, or inertia, which would push the water toward the bottom of the globe, are ignored. Furthermore, there is no evidence of deformation of the balloon, which, being an elastic element, should be affected by the forces." | Example: "Initial prompt: Generate a physically accurate image of a child swinging a water balloon on a string in a horizontal circle. The string should angle slightly downward due to gravity, the balloon should appear elongated due to centripetal force, and the water should spray tangentially due to inertia.' Follow-up 1: The balloon is still too spherical—due to the centripetal effect, the water inside should be pushed toward the outer wall, making the balloon more elongated in the outward direction. Follow-up 2: The child should be leaning back slightly because the tension required for circular motion ($T = mv^2/r$) would pull them toward the balloon—they need to brace against this force." |
| Part 3 | Example: "Create an FBD for the picture; additional prompt: Yes, create a labeled FBD sketch" | Example: "Draw the correct FBD of the water balloon being swung in a horizontal circle." | Example: "Initial: can you make a free body diagram of the forces in the image acting on the water balloon as it is being swung in a circle by the child. Follow-up: shouldn't there also be a force of gravity? and isn't the centripetal force supposed to point inwards?" | Example: "could you draw a free body diagrom of the water balloon being swung in a horizontal circle as the baby is swinging it in the picture. Include gravity and of the likes. Also could u make it into a square. Follow-up: Further prompt: Make sure to include tension, angle from vertical, centripetal acceleration that should be wards the center of the circle and gravity please. Also, the balloon is shown to be in the middle of the air so there shouldn't be normal forces or friction." | Example: "Initial prompt: Draw a correct free body diagram of a water balloon being swung in a horizontal circle. The string is angled downward from the hand to the balloon. Show and label the forces acting on the balloon: Tension T along the string toward the hand (toward the center), Weight mg downward. Mark the angle α between the string and the vertical. Indicate radial (toward the center) and vertical directions. Use the relations: $T \cdot \cos(\alpha) + Fb - mg = 0$ and $T \cdot \sin(\alpha) = m \cdot v^2/r$. Follow-up 1: Point the tension (T) vector strictly toward the center of the circle. Remove the separate centripetal force arrow and instead show it as $T\sin(\alpha) = mv^2/r$. Make sure the angle α is measured from the vertical line. Follow-up 2: Add the vertical equilibrium equation $T\cos\alpha + Fb - mg = 0$. Align the buoyant force (Fb) directly opposite to mg—both vertical." |

*Note.* **No physics-driven interaction (0):** Prompt contains no physics content; generic task request only. No physics-driven follow-up at any point in the interaction. **Minimal physics framing (1):** Physics topic mentioned in prompt but without specific physical conditions or parameters; no physics-driven follow-up. Student showed basic awareness of the disciplinary context but did not use physics knowledge to direct the AI interaction. *Part 2 note:* a prompt that simply references previously identified errors without restating specific physical conditions or justifications scores 1, not 2. *Part 3 note:* repeatedly asking AI to find and fix its own errors without student-supplied physics reasoning also

scores 1, regardless of number of iterations. **Adequate physics framing (2):** At least one specific physical condition with justification appears somewhere in the interaction. Path A: strong initial prompt specifying at least one physical condition, AI output sufficiently accurate, no corrective follow-up needed. Path B: weak initial prompt followed by at least one physics-driven follow-up directing the interaction toward disciplinary accuracy. **Proficient physics-driven interaction (3):** Multiple specific physical conditions appear in the interaction AND at least one physics-driven follow-up is present; OR initial prompt specifies multiple physical conditions and AI output was sufficiently accurate that no corrective follow-up was needed. Student actively controlled the interaction using disciplinary knowledge through structured multi-condition framing, with or without targeted iteration. **Advanced physics-driven interaction (4):** Prompt structures the interaction around multiple disciplinary goals AND multiple physics-driven follow-up iterations are present. Student demonstrated both proactive disciplinary framing and sustained iterative refinement—the full cycle of physics-driven AI interaction. *Task-specific physical conditions for Part 2 (image generation) include: balloon elongation or deformation due to centripetal force; string angled downward due to gravity; child leaning back due to tension required for circular motion; water spraying tangentially due to inertia; water distributed toward outer edge of balloon due to centripetal effect; irregular water stream due to pressure decay. Task-specific physical conditions for Part 3 (FBD generation) include: tension T along string toward pivot/center; weight mg downward; no normal force or friction since balloon is airborne; centripetal force is not an independent force but the horizontal component of tension ($T \sin\theta = mv^2/r$); angle θ measured from vertical.*

## Table 4. Scoring Rubrics for Physics-Grounded Evaluation (D2): Water Balloon Task

| Indicator | Scoring Rubrics | | | |
|---|---|---|---|---|
| | 0<br>No evaluation or inaccurate evaluation | 1<br>Surface-level evaluation | 2<br>Physics-grounded evaluation | 3<br>Modeling-level evaluation |
| Part 1 | Example: "I did not find anything it missed." | Example: "I did not see any inconsistencies or inaccuracies with GenAI, but they did not mention that the string would probably sag." | Example: "The 'steady' stream of water flowing out of the water balloon is impossible—water pressure inside balloon drops when holes open, outflow rate should decay exponentially due to Bernoulli principle. The stream shouldn't be consistent. Water should be parabolic due to Newton's first law. Also, the kid should be leaning back a little bit due to inertia—tension force mv²/r is needed but the kid doesn't display any strain or shift in balance." | Example: "String droops while motion is 'horizontal'—for horizontal circular motion, the string must be taut and horizontal. A droop implies unbalanced forces. Balloon maintains perfect shape—a real water balloon would deform or burst due to the outward centrifugal tension forces." |
| Part 3 | Example: "For the next prompts, I asked it to identify any mistakes about the image it has generated and to create a new one using matplotlib while fixing the mistakes in the previous image. I repeated this 4 times." | Example: "The image in the diagram is unclear, there are errors in the organization, and there are even forces that do not make sense." | Example: "The water balloon is not on the surface so there shouldn't be a normal force or friction. The tension along the string is at an angle and is not purely horizontal. The weight should be vertically downward, not balanced by tension or normal forces. Main dynamic is the centripetal acceleration towards the circle's center and not resisted by friction. Tension should be pointing inward." | Example: "It showed centripetal force as a separate physical force, which is incorrect since centripetal force is the horizontal component of tension. Follow-up: Remove the separate centripetal force arrow and instead show it as Tsin(α) = mv²/r. Make sure the angle α is measured from the vertical line. Add the vertical equilibrium equation Tcosα + Fb − mg = 0." |

***Note.*** **For Part 2: No evaluation or inaccurate evaluation (0):** Student provides no independent evaluation; accepts AI's analysis as complete without any physical assessment. OR student's independent judgment is inaccurate—accepts AI's incorrect claims without questioning, or incorrectly challenges AI's correct identification of expert key errors. No engagement with expert-defined key errors. **Surface-level evaluation (1):** Student shows some independent thinking but without physical justification. Includes: agreeing with AI's correct identification of expert key errors without adding reasoning; or identifying an additional error beyond AI's list but without connecting it to a physics principle. Student engages with the part but does not demonstrate physics-grounded reasoning. **Physics-grounded evaluation (2):** Student accurately judges AI's analysis with at least one physics-grounded addition or correction. This includes: correctly identifying at least one error with physical justification, even if the error is not among the expert-defined key errors but is physically valid; OR correctly challenging an AI claim with a disciplinary principle. Judgment is connected to at least one physics law or principle. **Modeling-level evaluation (3):** Student evaluates AI's analysis at the level of how AI has modeled the physical phenomenon—questioning AI's representational choices or the validity of AI's physical reasoning, not just adding omissions. Justification references disciplinary principles and engages with at least one expert key error. May include: identifying that AI's steady water stream assumption violates Bernoulli's principle; questioning the validity of AI's reasoning about gravity's effect; or identifying that AI ignored centripetal effects on internal water distribution. **For Part 3: No evaluation or inaccurate evaluation (0):** Student provides no independent evaluation of the FBD; accepts AI output without any physical assessment. Includes cases where error-finding is fully delegated back to AI rather than based on student's own physical understanding. AI FBD contained key errors but student did not engage with them. **Surface-level evaluation (1):** Student notices errors in the FBD or accepts a correct output, but without physical justification. Includes: identifying that something looks wrong without connecting it to a physics principle; vaguely noting missing or extra forces without explaining why they violate physics. Student shows independent engagement but reasoning is not grounded in disciplinary principles. **Physics-grounded evaluation (2):** Student accurately judges the AI FBD with at least one physics-grounded addition or correction. This includes: correctly identifying at least one error with physical justification, even if the error is not among the expert-defined key errors but is physically valid; OR correctly accepting an accurate FBD and demonstrating understanding of why it is correct. Judgment is connected to at least one physics law or principle. **Modeling-level evaluation (3):** Student identifies at least one expert-defined key error with physical justification. Justification references disciplinary principles and connects the identified error to a specific physics law or principle. For part 3, the expert-defined key error is: centripetal force drawn as a separate arrow is incorrect—it is the horizontal component of tension ($T \sin\theta = mv^2/r$), not an independent force.

## 5. Discussion

This study responded to two interrelated challenges in STEM assessment under GenAI: a validity challenge arising from GenAI's capacity to substitute for student cognition in conventional task formats, and a construct challenge arising from the emergence of disciplinary AI interaction as a competency for which few operationalizable assessment frameworks currently exist. To address both, this study proposed an ECD-based framework that analyzes GenAI's impact across the student model, evidence model, and task model, and used this analysis to articulate three principled governance stances—Restrict, Scaffold, and Require—each grounded in a validity argument rather than a blanket policy about permissible use. To illustrate the Require stance, two task designs were deployed in an introductory physics course. Across both tasks, two dimensions of disciplinary AI interaction competency—Physics-Driven AI Interaction and Physics-Grounded Evaluation—were shown to be observable in student process artifacts and scorable using rubrics calibrated against expert knowledge, demonstrating that disciplinary AI interaction competency is observable and scorable in student process artifacts using rubrics calibrated against expert knowledge.

### *5.1. The ECD-based framework as a validity-grounded alternative to existing AI use policies*

The ECD-based framework proposed in this study repositions GenAI governance within the assessment argument rather than treating it as an external policy matter. Rather than asking what level of AI use should be permitted, the framework asks a prior question: how does AI integration reshape the inferential chain from observable student work products to targeted constructs? By analyzing this question through the three ECD models—the student model, the evidence model, and the task model—the framework generates governance decisions that are grounded in validity logic rather than institutional preference. A Restrict decision is warranted not because AI use is generally undesirable, but because in a specific assessment context, AI can substitute for the reasoning the construct targets, breaking the inferential link between student work and targeted proficiency. A Scaffold decision is warranted when AI can be guardrailed so that it reduces peripheral cognitive demands without revealing the target solution, preserving the evidential integrity of student responses in the same way that a calculator preserves the evidential integrity of a mathematical proof (Kapoor et al., 2025). A Require decision is warranted when AI interaction competency is itself the target construct, making AI use not a threat to validity but a constitutive element of the assessment argument.

The two tasks presented in this study illustrate how this validity logic translates into concrete assessment design. Both tasks satisfy the joint conditions for Require governance: AI interaction competency is the target construct, process artifacts make student cognitive contribution observable and separable from AI-generated output, and the accuracy-ceiling criterion confirms that student disciplinary input is necessary to reach the highest performance levels. This last condition—that GenAI consistently missed at least one expert-anchored error in both tasks—is particularly important for distinguishing Require from what might superficially resemble AI collaboration in existing frameworks (Perkins et al., 2024). In an AI collaboration task as conventionally defined, students may simply accept and lightly edit AI output; in a Require task as defined here, the task is specifically designed so that uncritical acceptance of AI output is detectable and produces a lower-quality artifact than disciplinary engagement does. The distinguishability condition—that student artifacts must make cognitive contribution observable and separable from AI output—is what operationalizes this distinction and what existing use-policy frameworks do not provide a principled basis for specifying. By grounding this condition in the evidence model rather than leaving it implicit, the ECD-based framework offers assessment designers a validity-anchored criterion for determining not just whether AI use is permitted, but whether the resulting student artifacts can support defensible inferences about targeted competencies (Kane, 2013; Mislevy et al., 2003).

### *5.2. GenAI as Scaffold and Interlocutor: Design Principles and Open Questions*

A recurring assumption in current discussions of GenAI governance is that AI's primary role in assessment is as a threat—an answer generator that students may use to outsource cognitive work, undermining the validity of inferences drawn from their performance (Zhao et al., 2024). The framework proposed in this study suggests that this assumption captures only one of several roles GenAI can play in assessment contexts, and that recognizing the distinction between GenAI as scaffold and GenAI as interlocutor has direct implications for how assessment tasks should be designed, what evidence should be collected, and what inferences can be defensibly drawn from student performance.

When GenAI functions as a scaffold, it is permitted to handle peripheral cognitive demands—such as retrieving factual information or providing worked examples—while the student continues to demonstrate the core disciplinary reasoning the assessment targets. This role is analogous to a calculator in a mathematical proof or a reference manual in experimental design: the tool reduces irrelevant cognitive load without substituting for the reasoning the construct requires (Liffiton et al., 2023). In a Scaffold governance stance, the key design challenge is specifying the boundary between what GenAI may and may not do—guardrailing AI support so that it does not cross from peripheral assistance into target-construct substitution. A modified version of the Water Balloon task illustrates this role concretely: if students were provided with a pre-generated AI image and FBD and asked only to identify errors and evaluate the output—removing the prompting phase entirely—the governance stance would shift from Require to Scaffold. The target construct would shift from AI interaction competency to physics evaluation competency, Physics-Driven AI Interaction would no longer be elicited or scored, and GenAI would function as a delivery mechanism for eliciting evidence about a non-AI construct rather than as an interlocutor whose outputs students must actively shape. The evidential integrity of Physics-Grounded Evaluation would be preserved, but the task would no longer assess whether students can use disciplinary knowledge to drive AI interaction—only whether they can evaluate its outputs.

When GenAI functions as an interlocutor, the governance logic shifts fundamentally. Rather than being guardrailed to prevent substitution, AI is positioned as a dialogue partner whose outputs students must actively shape, evaluate, and correct using disciplinary knowledge. In this role, AI interaction competency is itself the target construct, and the task is designed not despite the presence of GenAI but around it. Both tasks presented in this study instantiate this role: the Remora task requires students to guide AI toward a biologically accurate FBD that AI cannot generate without disciplinary input, and the Water Balloon task requires students to identify physically impossible features that AI consistently misses and to correct AI-generated diagrams using Newtonian mechanics. The key design condition that makes this role viable—rather than collapsing into AI collaboration as conventionally defined (Perkins et al., 2024)—is the accuracy-ceiling criterion: GenAI must consistently miss at least one expert-anchored error, ensuring that student disciplinary input is necessary to reach the highest performance levels and that uncritical acceptance of AI output is both detectable and penalized in scoring. In both tasks, the expert answer key confirmed that this condition was satisfied, empirically validating the Require stance and establishing that the process artifacts students produced—prompts, critiques, and reflective commentaries—provide defensible evidence of disciplinary AI interaction competency that cannot be outsourced to AI itself (Li et al., 2025; Mao et al., 2024).

The boundary between these two roles is not fixed by the presence or absence of AI in the task, but by whether the interaction with AI is itself the inferential target or merely a delivery mechanism for eliciting evidence about a non-AI construct. This distinction has practical consequences for task design: a task that appears to require AI interaction may be scaffolding it if the AI outputs are sufficiently accurate that students can reach the highest performance levels without deploying disciplinary knowledge. Conversely, a task that appears to scaffold AI use may be requiring it if the design inadvertently makes AI interaction quality the primary source of variance in student performance. The ECD-based framework provides a principled basis for navigating this boundary—by anchoring the governance decision in the student model (what construct is targeted), the evidence model (what artifacts make student reasoning observable), and the task model (what design features ensure the accuracy-ceiling condition is satisfied) — rather than relying on surface features of the task or the degree to which AI is visibly present (Kane, 2013; Mislevy et al., 2003).

While the distinction between GenAI as scaffold and GenAI as interlocutor provides a principled basis for task design decisions, it also surfaces a set of empirical questions that the ECD framework can specify but cannot yet answer. The first concerns the actual effectiveness of Scaffold governance. The framework assumes that guardrailed AI support reduces peripheral cognitive demands without altering the reasoning processes that the assessment targets—but this assumption has not been empirically verified. It remains an open question whether students who interact with guardrailed AI engage in qualitatively different disciplinary reasoning than students who work without AI support, or whether even bounded AI involvement subtly reshapes the cognitive pathway through which students arrive at their responses. Establishing the validity of Scaffold governance requires not only that the guardrail boundary be correctly specified, but that empirical evidence confirm that student reasoning within that boundary is substantively equivalent to the unassisted reasoning the construct was originally designed to measure (Kane, 2013).

The second concerns the dynamic relationship between GenAI capability and the scaffold-interlocutor boundary. As frontier models continue to improve, tasks that currently satisfy the accuracy-ceiling criterion—and therefore warrant Require governance—may eventually be solvable by GenAI without student disciplinary input, collapsing the Require stance into something closer to AI collaboration or even full outsourcing (Jian et al., 2025; Kortemeyer, 2023). Similarly, tasks that currently warrant Scaffold governance because GenAI's substitution capacity is limited may eventually require Restrict governance as that capacity expands. The framework provides decision criteria for navigating these shifts, but the empirical monitoring required to detect when a task has crossed a governance boundary remains an unresolved methodological challenge—one that will require ongoing collaboration between assessment designers, disciplinary experts, and researchers tracking the frontier of GenAI capability.